\documentclass{mem}
\usepackage{natbib}
\usepackage{txfonts}
\usepackage{balance}
\usepackage{graphicx}
\usepackage[a4paper,breaklinks]{hyperref}
\begin{document}

\title{
Revisiting the Statistics of X-ray Flares in Gamma-ray Bursts}

   \subtitle{}

\author{
Y. \,Wang\inst{1,2} 
\and Y. \, Aimuratov\inst{1,2,4}
\and R. \, Moradi\inst{1,2}
\and M. \, Peresano\inst{3}
\and R. \, Ruffini\inst{1,2,4,5}
\and S. \, Shakeri\inst{2,6}
          }

\institute{
ICRA and Dipartimento di Fisica, Sapienza Universit\`a di Roma, P.le Aldo Moro 5, 00185 Rome, Italy
\and
ICRANet, P.zza della Repubblica 10, 65122 Pescara, Italy
\and
Universit\`a degli Studi di Udine, via delle Scienze 206, 33100 Udine, Italy
\and
Universit\'e de Nice Sophia Antipolis, CEDEX 2, Grand Ch\^{a}teau Parc Valrose, Nice, France
\and
ICRANet-Rio, Centro Brasileiro de Pesquisas F\'isicas, Rua Dr. Xavier Sigaud 150, 22290--180 Rio de Janeiro, Brazil
\and
School of Astronomy, Institute for Research in Fundamental Sciences (IPM), P.O. Box 19395-5531, Tehran, Iran
\\
\email{yu.wang@icranet.org}
}

\authorrunning{WANG}

\titlerunning{X-ray Flares}

\abstract{
The statistics of X-ray flares in the afterglow of gamma-ray bursts (GRBs) have been studied extensively without considering 
the possible different origins of each flare. By satisfying six observational criteria, we find a sample composed of $16$ long GRBs 
observed by \textit{Swift} satellite may share a same origin. By applying the Markov chain Monte Carlo iteration and the 
machine learning algorithms (locally weighted regression and Gaussian process regression), impressively, the flares in 
these GRBs show strong correlations with the energy released in the prompt emission. These correlations were never 
discovered in previous papers, and they could not be well explained by previous models. These correlations imply that 
the prompt emission and the X-ray flare are not independent, they may be originated following a same sequence. The new 
THESUS satellite will provide us a larger sample and more detailed spectra to refine the results we obtained in this 
article.

}
\maketitle{}

\section[Introduction]{Introduction}
\label{sec:intro}

GRBs are complex authentic astrophysical laboratories encompassing different families, originating in different progenitors, and each characterized by different physical and astrophysical regimes. \citet{1992AIPC..265..304D, 1993ApJ...413L.101K} classified GRBs into short or long classes due to the duration of the prompt emission, \citet{2015IJMPA..3045023R, 2016ApJ...832..136R} proposed several families of GRBs based on their different configurations of binary systems.

The observed time sequence of a typical long GRB starts from the trigger time given by an instrument on-board of a satellite, often \textit{Swift}-BAT or \textit{Fermi}-GBM. Later the instrument registers a very intense emission, prompt emission, lasting tens of seconds, and constituted by one or more pulses. This emission contains the majority of the GRB energy in hard X-ray and gamma-ray. Then comes the afterglow, observed by \textit{Swift}-XRT in the soft X-ray band, and several telescopes in the optical and radio bands. Afterglow starts, following the tail of prompt emission, with one or more flares superposing, these flares last some hundred of seconds. Then a shallow decay or a plateau during the time range $10^2$ - $10^6$ s. Later the light curve decays again with a power-law index $\sim -1.2$  \citep{2006ApJ...642..354Z}. During the prompt emission and the afterglow, a few GRBs observed high energy GeV photons by Fermi-LAT \citep{2013ApJS..209...11A}.  The observation ceases commonly in a few days or months. In this paper, we concentrate on the X-ray flares in the early afterglow and their relations with the aforementioned processes.

%This general picture conforms well with most of the long duration energetic GRBs ($T90 > 2$~s, isotropic energy $E_{iso} > 10^{52}$~erg), while short GRBs and low energy long ones have variations \citep{2016ApJ...832..136R}.

Flares in the afterglow appear approximately in half of the observed GRBs \citep{2013NatPh...9..465W}, some of which have more than one flare.  Flares mostly arise in the early afterglow ($ < 1000 s$), while only a few are observed in the very late time ($>10^4 s$). \citet{2007ApJ...671.1903C} fitted more than $60$ X-ray flares of short and long GRBs together, most of them have no detection of redshift, they found the duration of flares and their peak time are faintly correlated, and that the subgroup of them occurring earlier than $1000~s$ can hardly form a correlation. \citet{2008A&A...487..533C} classified flares into early and late groups, and their result confirms \citet{2007ApJ...671.1903C}.

In this paper\footnote{The data analysis of this work was performed during the Swift data reduction seminar 
given by Y.~Wang in ICRANet under the supervision of R. Ruffini. Y.~Aimuratov, R.~Moradi, M.~Peresano and S.~Shakeri 
attended the seminar and all contributed to this work. Related material has been published on \url{http://grb.physical.reviews/en/latest/XRT.html}}, 
our approach is not to improve on the existing statistical analysis containing as more flares as possible. Instead we are open to a more selective vision that very distinct patterns of flares can be observed, and consequently the underlying mechanisms causing similar morphologies can be identified. We consequently select a group of flares sharing some similarities, and fulfilling our criteria (Section 2). We analyze the flares and general features of these GRBs. Our sample shows clear new correlations, some of which are not prominent if adopting all the flares (Section 3). We discuss that the previous models lacks the capacity of interpreting our findings (Section 4).

\section{Sample Preparation} \label{sec:sample}

 We adopt a machine learning algorithm named locally weighted regression \footnote{\url{https://github.com/YWangScience/AstroNeuron}}, similar sample is obtained from \citet{2013ApJ...774....2S} which applied the Bayesian Information method. Additionally, we filter the GRBs again by applying 6 model independent criteria. Our purpose of these criteria follows:

\begin{enumerate}

\item[] Most previous papers adopt all the GRBs, even without known redshift, this is inappropriate to analyze the physical origins, especially for high redshift GRBs.

\item[] Most previous papers analyze all the x-ray flares together, which ignores possible findings and correlations existing in specific subclasses of GRBs.

\end{enumerate}

We intend to collect a sample of flares possibly from a same mechanism. We start from the GRBs with known redshift, since we care most the intrinsic mechanism. Then we make two hypotheses to be justified, and our criteria are based on the hypotheses.

\textit{Hypothesis 1}: Flares with distinct observed patterns could be produced by various mechanisms.
So we select select flares sharing similar morphologies, which brings two criteria:

\begin{enumerate}

\item[] Criterion 1: Flares in the long GRBs.

\item[] Criterion 2: Flares occurring before the plateau phase: early time flare.

\end{enumerate}

\textit{Hypothesis 2}: X-ray flares could be generated differently from the prompt spikes, which also brings two criteria:

\begin{enumerate}

\item[] Criterion 3: Exclude flares that other bands dominate the soft X-ray band.

\item[] Criterion 4: Exclude flares contaminated by the prompt emission.

\end{enumerate}

Also in order to do reliable data analysis, we consider 

\begin{enumerate}

\item[] Criterion 5: Obvious flare: luminosity at the peak of the flare must be more than double of the underlying light curve, and the signal to noise ratio (SNR) in the flare $>$10.

\item[] Criterion 6: Swift X-ray light-curve is complete till $10^4$~s in the cosmological rest-frame.

\end{enumerate}

\begin{figure}
\centering
\includegraphics[width=1\hsize,clip]{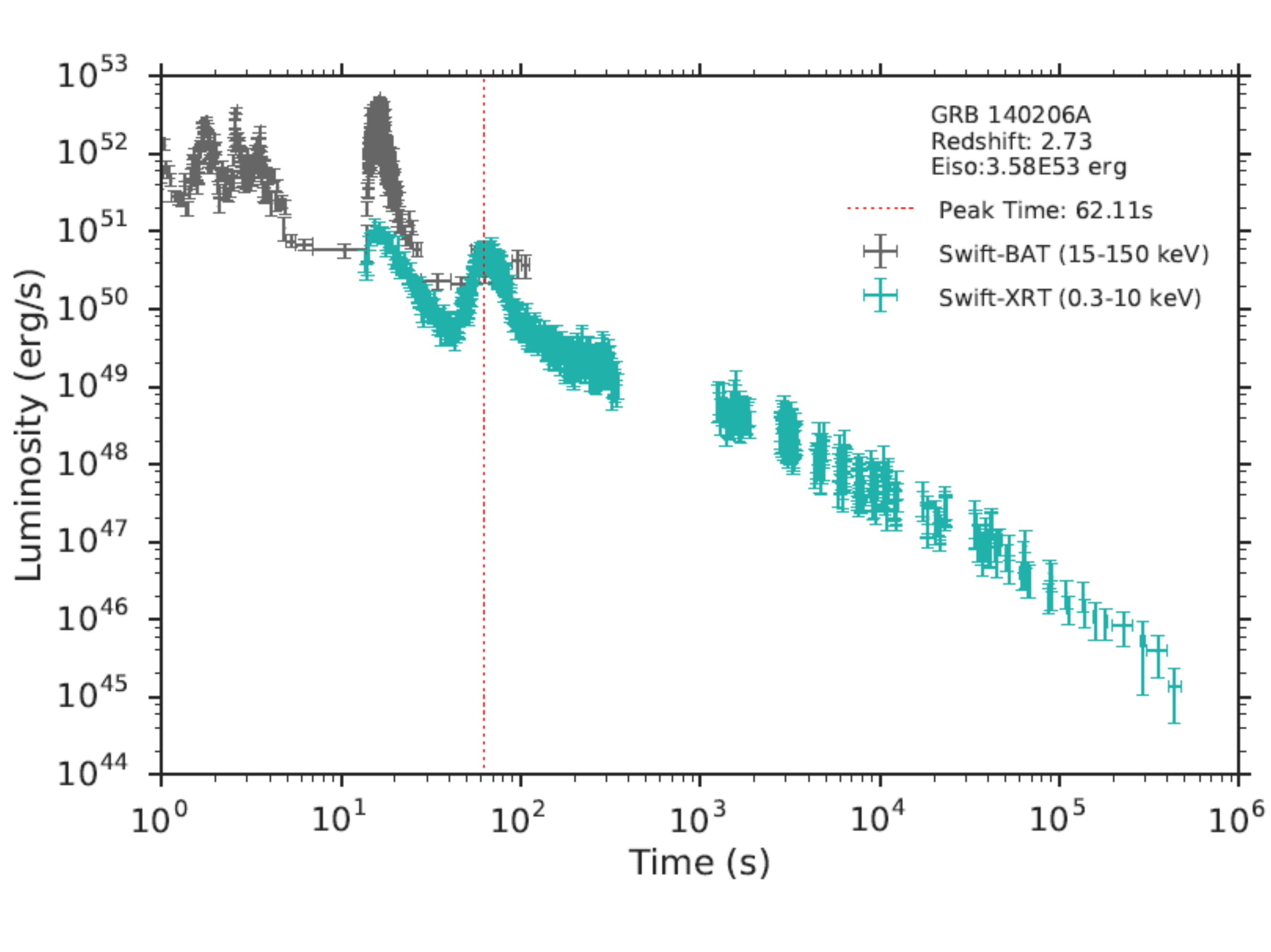}
\caption{\textbf{Light-curve of GRB 140206A:} This GRB was detected by \textit{Swift} \citep{2014GCN..15784...1L}. The \textit{Swift}-BAT light curve shows a multi-peaked structure with roughly three main pulses \citep{2014GCN..15805...1S}. The redshift, as observed by NOT \citep{2014GCN..15800...1M} is $z=2.73$, and the isotropic energy is $E_{iso}=3.58\times 10^{53}$~erg. GRB 140206A has two flares in \textit{Swift}-XRT. A gamma-ray flare coincides with the last \textit{Swift}-BAT spike at $\sim 18$~s having a spectral of power-law index $-0.88 \pm 0.03$, most of the energy is contributed by the hard X-ray and the gamma-ray photons. The second flare at $62.11$ s with a power-law index $-1.73 \pm 0.06$ is the flare that we are interested in this paper, most of its energy is in the soft X-ray band. The light-curve is plotted in the cosmological rest-frame, k-correction is dually applied.}
\label{fig:140206A}
\end{figure}

Surveying $421$ Swift GRBs with redshift till the end of $2016$ year, we find $16$ GRBs satisfying all the criteria. Among them, $7$ GRBs show a single X-ray flare. The other $9$ GRBs contain two flares, generally we exclude the first one by Criterion 3. This sample covers a wide range of redshift. GRB 070318 is the closest one with redshift $z=0.84$, GRB 090516A with redshift $z=4.11$ is the farthest one. The isotropic energy $E_{iso}$ also spreads a wide range: 5 GRBs have $E_{iso}$ of the order of $10^{52}$~erg, 9 GRBs have $E_{iso}$ the order of $10^{53}$~erg, and 2 GRBs have extremely high $E_{iso}$ $E_{iso} > 10^{54}$~erg. Therefore, we consider this sample is well-constructed. We give an example of the a selected GRB 140206A in figure \ref{fig:140206A} \footnote{The entire luminosity light-curves of the 16 GRBs are shown in \citet{2018ApJ...852...53R}}.

%\footnote{Technical details concerning selecting GRBs using machine learning algorithm and the data analysis using HEASoft are explained in \citet{2018ApJ...852...53R}}

\begin{table*}
\scriptsize
\centering
\begin{tabular}{ccccccc}
\hline\hline
GRB & z &  $E_{iso}$ (erg) & $t_p$ (s) & $L_p$ (erg/s) & $\Delta t$ (s) & $E_{f}$ (erg)\\
\hline
060204B & 2.3393 & $2.93(\pm0.60)\times10^{53}$  & $100.72\pm6.31  $ & $7.35(\pm2.05)\times10^{49}$  & $17.34\pm6.83 $ & $8.56(\pm0.82)\times 10^{50}$  \\
060607A & 3.082  & $2.14(\pm1.19)\times10^{53}$  & $66.04\pm4.98  $ & $2.28(\pm0.48)\times10^{50}$  & $18.91\pm3.84 $ & $3.33(\pm0.32)\times 10^{51}$  \\
070318  & 0.84   & $3.41(\pm2.14)\times10^{52}$  & $154.7\pm12.80 $ & $6.28(\pm1.30)\times10^{48}$  & $63.80\pm19.82$ & $3.17(\pm0.37)\times 10^{50}$  \\
080607  & 3.04   & $1.87(\pm0.11)\times10^{54}$  & $37.48\pm3.60  $ & $1.14(\pm0.27)\times10^{51}$  & $15.63\pm4.32 $ & $1.54(\pm0.24)\times 10^{52}$  \\
080805  & 1.51   & $7.16(\pm1.90)\times10^{52}$  & $48.41\pm5.46  $ & $4.66(\pm0.59)\times10^{49}$  & $27.56\pm9.33 $ & $9.68(\pm1.24)\times 10^{50}$  \\
080810  & 3.35   & $5.00(\pm0.44)\times10^{53}$  & $51.03\pm6.49  $ & $1.85(\pm0.53)\times10^{50}$  & $12.38\pm4.00 $ & $1.80(\pm0.17)\times 10^{51}$  \\
081008  & 1.967  & $1.35(\pm0.66)\times10^{53}$  & $102.24\pm5.66 $ & $1.36(\pm0.33)\times10^{50}$  & $18.24\pm3.63 $ & $1.93(\pm0.16)\times 10^{51}$  \\
081210  & 2.0631 & $1.56(\pm0.54)\times10^{53}$  & $127.59\pm13.68 $ & $2.23(\pm0.21)\times10^{49}$  & $49.05\pm6.49$ & $8.86(\pm0.54)\times 10^{50}$  \\
090516A & 4.109  & $9.96(\pm1.67)\times10^{53}$  & $80.75\pm2.20  $ & $9.10(\pm2.26)\times10^{50}$  & $10.43\pm2.44 $ & $7.74(\pm0.63)\times 10^{51}$  \\
090812  & 2.452  & $4.40(\pm0.65)\times10^{53}$  & $77.43\pm16.6  $ & $3.13(\pm1.38)\times10^{50}$  & $17.98\pm4.51 $ & $5.18(\pm0.61)\times 10^{51}$  \\
131030A & 1.293  & $3.00(\pm0.20)\times10^{53}$  & $49.55\pm7.88  $ & $6.63(\pm1.12)\times10^{50}$  & $33.73\pm6.55 $ & $3.15(\pm0.57)\times 10^{52}$  \\
140206A & 2.73  & $3.58(\pm0.79)\times10^{53}$  & $62.11\pm12.26 $ & $4.62(\pm0.99)\times10^{50}$  & $26.54\pm4.31 $ & $1.04(\pm0.59)\times 10^{51}$  \\
140301A  & 1.416 & $9.50(\pm1.75)\times10^{51}$  & $276.56\pm15.50  $ & $5.14(\pm1.84)\times10^{48}$  & $64.52\pm10.94 $ & $3.08(\pm0.22)\times 10^{50}$ \\
140419A & 3.956  & $1.85(\pm0.77)\times10^{54}$  & $41.00\pm4.68  $ & $6.23(\pm1.45)\times10^{50}$  & $14.03\pm5.74 $ & $7.22(\pm0.88)\times 10^{51}$ \\
141221A & 1.47  & $6.99(\pm1.98)\times10^{52}$  & $140.38\pm5.64 $ & $2.60(\pm0.64)\times10^{49}$  & $38.34\pm9.26 $ & $7.70(\pm0.78)\times 10^{50}$ \\
151027A & 0.81   & $3.94(\pm1.33)\times10^{52}$  & $183.79\pm16.43$ & $7.10(\pm1.75)\times10^{48}$  & $163.5\pm30.39$ & $4.39(\pm2.91)\times 10^{51}$ \\
\hline
\end{tabular}
\caption{GRB sample properties of the prompt and flare phases in the cosmological rest frame. This table contains: the redshift $z$, the isotropic energy $E_{iso}$, the flare peak time $t_p$, the flare peak luminosity $L_p$, the flare duration of which the starting and ending time correspond to half of the peak luminosity $\Delta t$, the flare energy $E_{f}$ within the time interval $\Delta t$.} 
\label{tab:grbList}
\end{table*}

\section{Statistical Correlation} \label{sec:correlation}

Physical mechanisms are reflected by observations, from which the most direct and obvious observables are energy and time. If our sample really reflects some physical processes, there shall exist some correlations between energy and time, or between themselves.

A GRB's energy is usually measured by the isotropic energy $E_{iso}$, which assumes that the prompt emission to be isotropic and is computed by integrating the prompt photons in the energy range from $1$ KeV to $10$ MeV \cite{2001AJ....121.2879B}. In our sample, \textit{Swift} has data for all the GRBs, \textit{Konus-Wind} observed GRB 080607, 080810, 090516A, 131030A, 140419A, 141221A and 151027A, while \textit{Fermi}  detected GRB 090516A, 140206, 141221A, 151027A (these 4 \textit{Fermi} GRBs have no prominent GeV emission with likelihood test statistic $TS < 20$). 

None of the satellite is able to cover the entire $1$ KeV to $10$ MeV energy band of $E_{iso}$, we need to fit the spectrum and find the best model, then extrapolate the integration of energy by the model. This method is relatively safe for GRBs observed by \textit{Fermi} and \textit{Konus-Wind}, but 8 GRBs in our sample have been observed only by \textit{Swift}, so we uniformly fit and extrapolate these 8 GRBs by power-law and cutoff power-law, then we take the average value as $E_{iso}$. In general, our priority of computing $E_{iso}$ is \textit{Fermi}, \textit{Konus-Wind}, then \textit{Swift}. The energy in the X-ray afterglow is computed in the cosmological rest-frame energy band from $0.3$~keV to $10$~keV. We smoothly fit the luminosity light-curve using locally weighted regression\footnote{\url{https://github.com/YWangScience/AstroNeuron}} which provides a sequence of power-law functions. The corresponding energy in a fixed time interval is obtained by summing up all of the integrals of the power-laws within it.  In order to take account for the expansion of the Universe, all our computation considers $k$-correction \citep{2001AJ....121.2879B}. To compute $E_{iso}$, the formula for $k$-correction varies depending on the best fitted model. For the light-curve of X-ray afterglow, $k$-correction is applied to each light-curve point with a power-law spectral index provided by \textit{Swift}-XRT online repository based on hardness ratio \citep{2007A&A...469..379E, 2009MNRAS.397.1177E}. Table. \ref{tab:grbList} lists the relevant energy value, also it lists the time information of prompt emission, X-ray flare and plateau.  The detection of flare's peak is realized by fitting the afterglow data points using also the locally weighted regression, which results in a smooth light-curve, the flare's peak is localized where the tangent of the light-curve has zero slope. The peak time $t_{peak}$ is defined as the time interval between the flare's peak and the trigger time, which is satellite dependent, all the sample we refer to \textit{Swift}-BAT. To extract the duration $\Delta t$, we work only with the most luminous part of the flare, taking the start time and the end time of the flare at the time when the luminosity is half of the one of the flare peak time.

\begin{figure}
\centering
\includegraphics[width=1\hsize,clip]{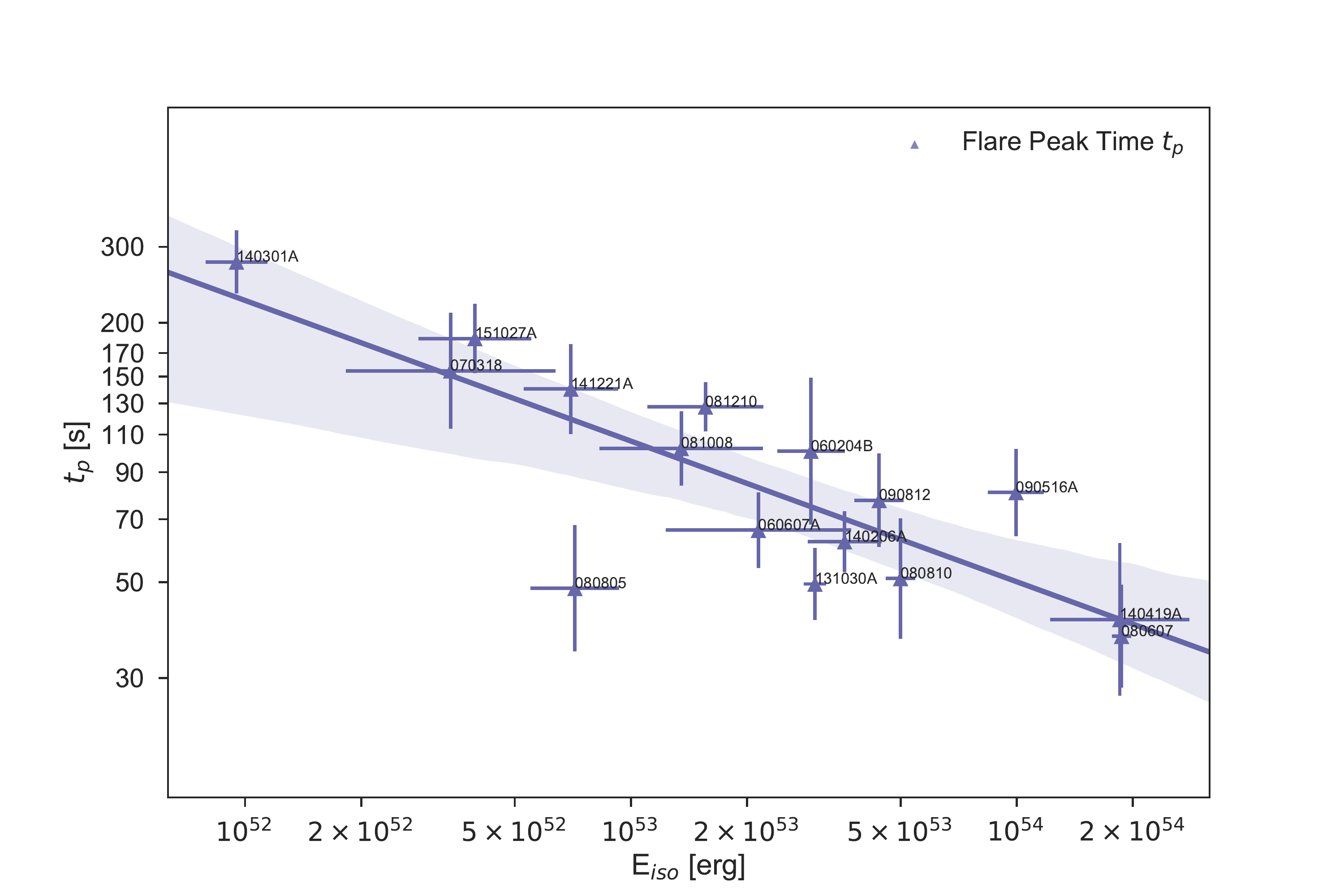}
\caption{Relation between $E_{iso}$ and $t_p$ fit by a power-law. The shaded area indicates the $95\%$ confidence level.}
\label{fig:EisoTflare}
\end{figure}

\begin{figure}
\centering
\includegraphics[width=1\hsize,clip]{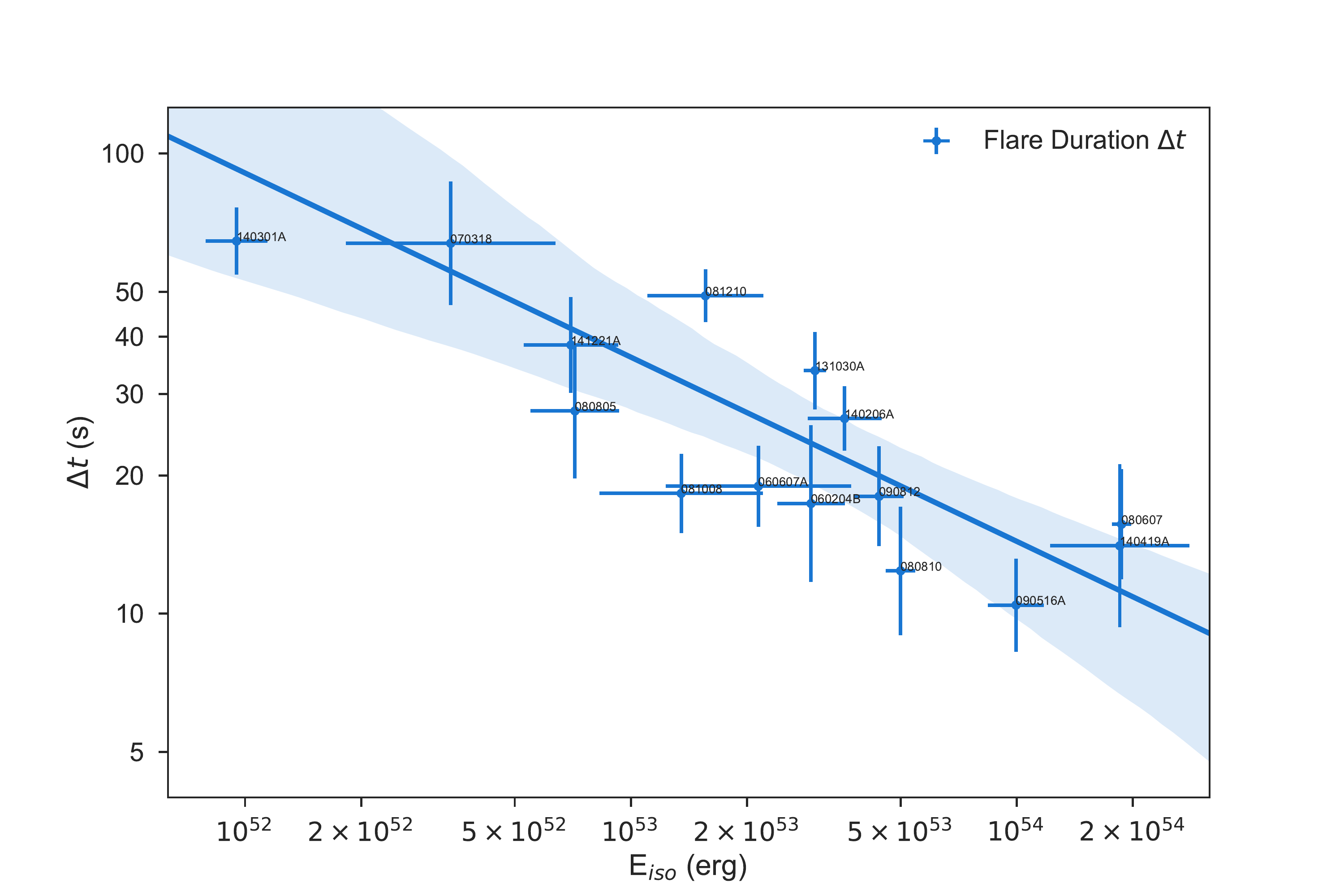}
\caption{Relation between $E_{iso}$ and $\Delta t$ fit by a power-law. The shaded area indicates the $95\%$ confidence level.}
\label{fig:EisoDeltaT}
\end{figure}

\begin{figure}
\centering
\includegraphics[width=1\hsize,clip]{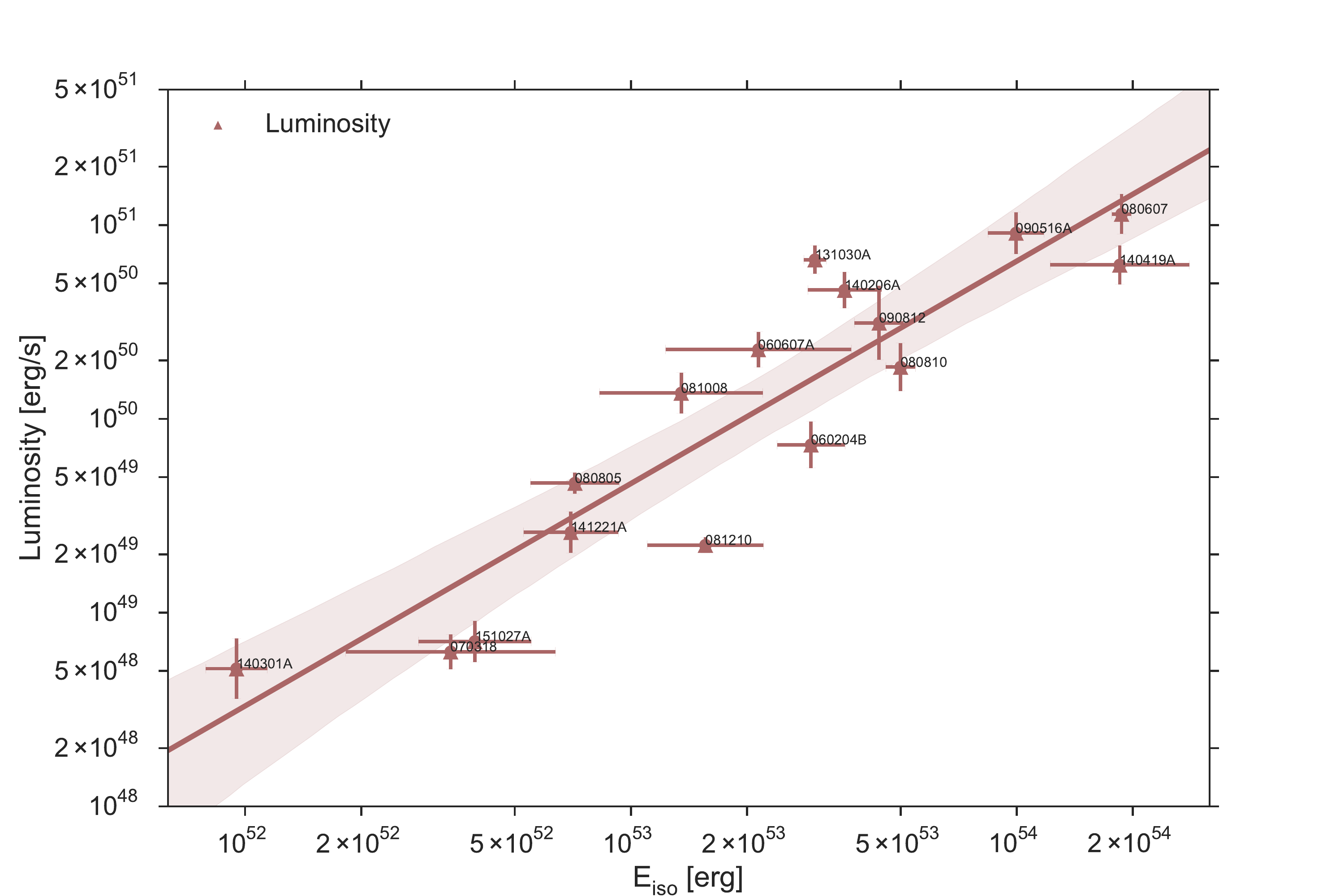}
\caption{Relation between $E_{iso}$ and $L_p$ fit by a power-law. The shaded area indicates the $95\%$ confidence level.}
\label{fig:EisoLuminosity}
\end{figure}

\begin{figure}
\centering
\includegraphics[width=1\hsize,clip]{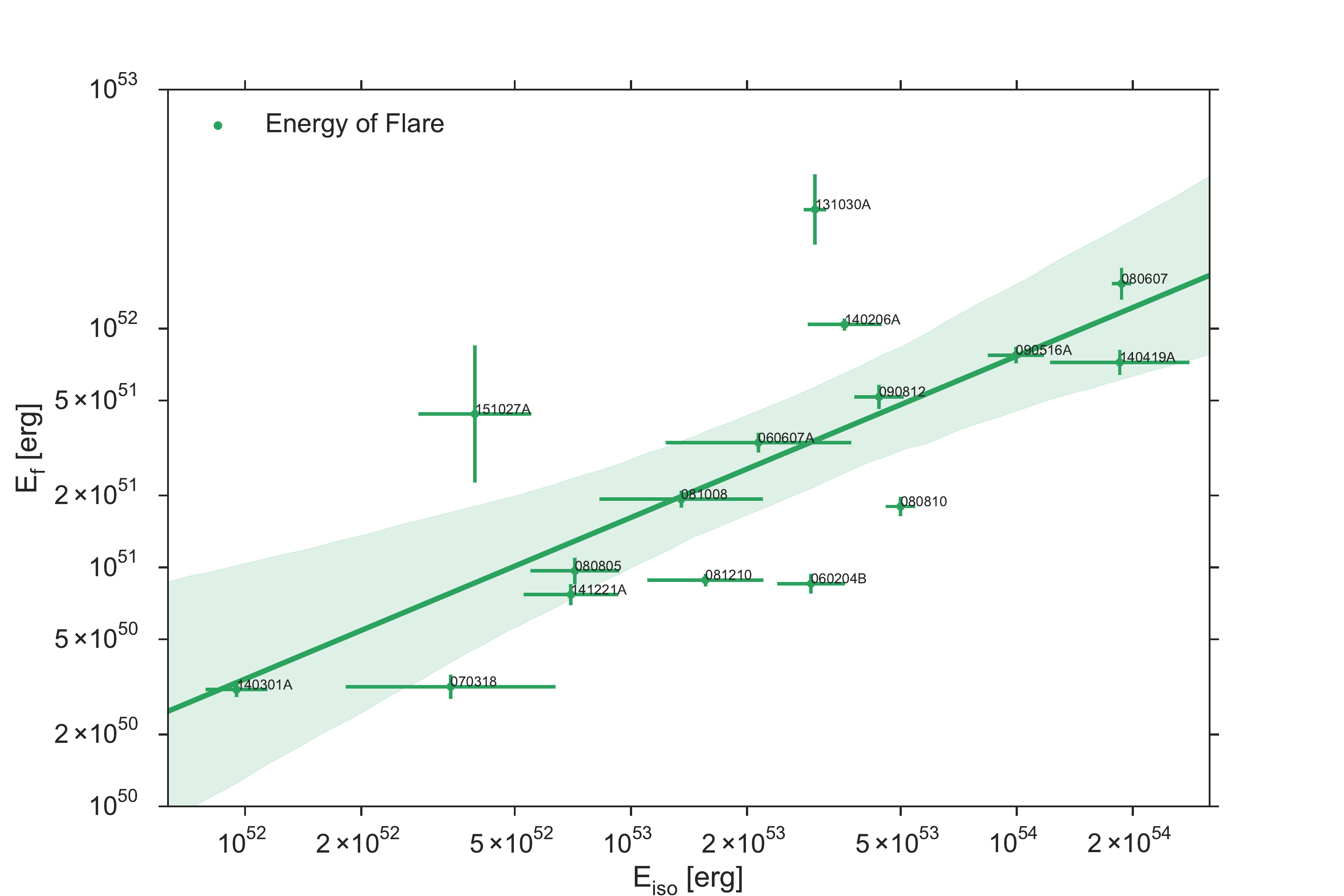}
\caption{Relation between $E_{iso}$ and $E_{f}$ fit by a power-law. The shaded area indicates the $95\%$ confidence level.}
\label{fig:EisoEflare}
\end{figure}

\begin{table}
\centering
\begin{tabular}{c@{\hskip 0.08in}c@{\hskip 0.08in}c@{\hskip 0.08in}}
\hline\hline
Correlation & Index & Coefficient \\
\hline
$E_{iso}-t_p$       & $-0.290(\pm0.010)$  & $-0.764(\pm 0.123)$  \\
$E_{iso}-\Delta t$  & $-0.461(\pm0.042)$  & $-0.760(\pm 0.138)$  \\
$E_{iso}-L_p$       & $1.186(\pm0.037)$   & $0.883(\pm 0.070)$  \\
$E_{iso}-E_f$       & $0.631(\pm0.117)$   & $0.699(\pm 0.145)$  \\
\hline
\end{tabular}
\caption{Power-law correlations among the quantities in Tab.~\ref{tab:grbList}. The values and uncertainties (at $1$--$\sigma$ confidence level) of the power-law index and of the correlation coefficient are obtained from $10^5$ MCMC iterations. All relations are highly correlated.}
\label{tab:correlation}
\end{table}

We then establish correlations between the above quantities characterizing the flare in the beginning of the afterglow with the isotropic energy of the prompt emission. We perform Markov chain Monte Carlo (MCMC) method and iterate $10^5$ times for having the best fit of the power-law and to obtain their correlation coefficients. The main results are summarized in figures ~\ref{fig:EisoTflare}--\ref{fig:EisoEflare} and in table~\ref{tab:correlation} \footnote{Codes are uploaded online \url{https://github.com/YWangScience/MCCC}}. We find that, for our sample that may present flares from a same origin, the peak time $t_p$, the duration of the flare $\Delta_t$, the peak luminosity $L_p$ and the total energy of flare $E_f$ are highly correlated with isotropic energy $E_{iso}$, all correlation coefficients are larger than $0.6$ (or smaller than $-0.6$).

\begin{figure}
\centering
\includegraphics[width=1.\hsize,clip]{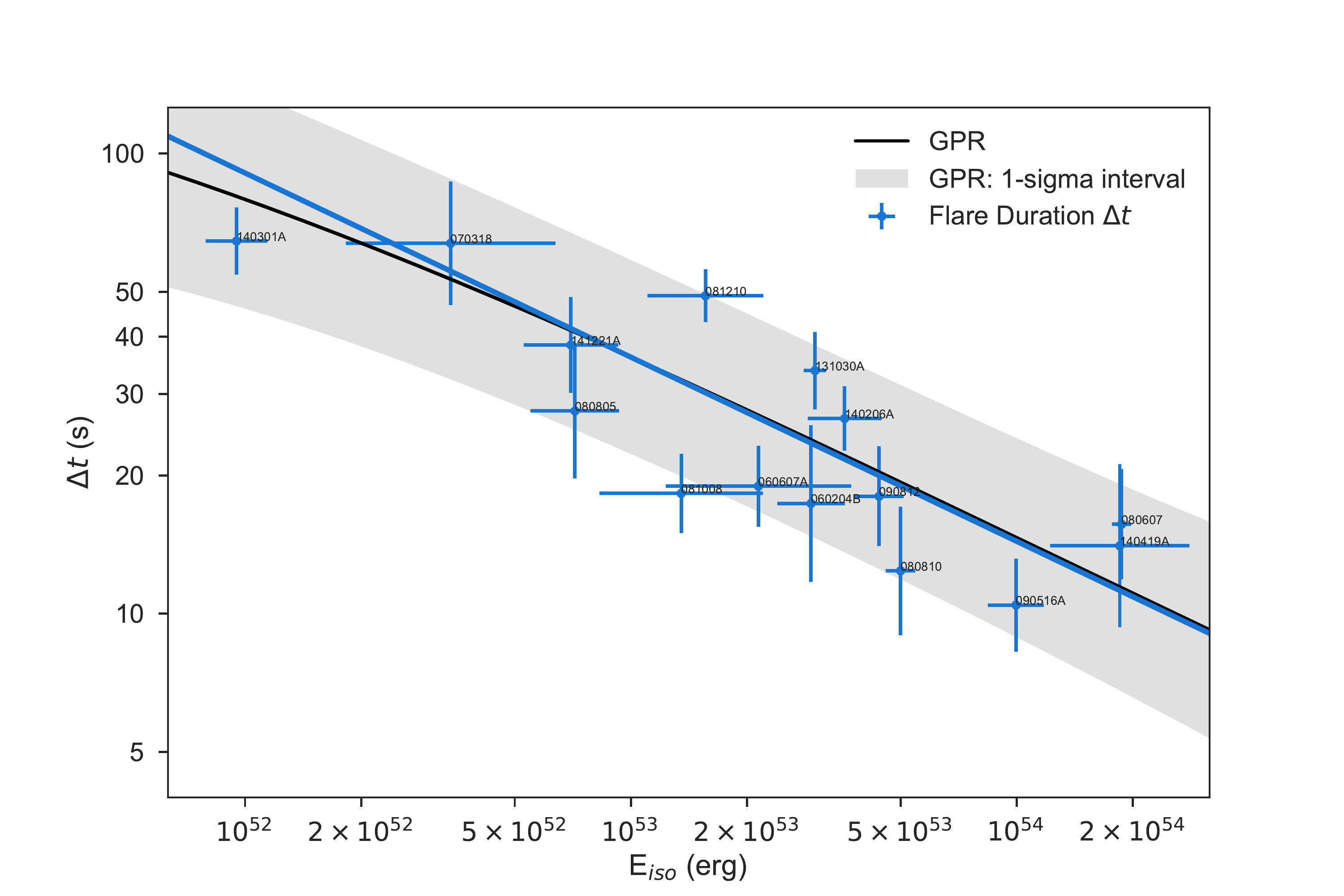}
\caption{The GPR fitted curve and its 1-sigma region in black and grey, and the power-law fitting by MCMC in blue (same as figure \ref{fig:EisoDeltaT}). The GPR curve almost coincides with the power-law line, which indicates the power-law assumption is appropriate.}
\label{fig:GPR}
\end{figure}

In order to verify that if the power-law well describes the correlations, we applied the Gaussian process regression (GPR), which is also an algorithm from machine learning, and it has been pioneeringly introduced to the area of cosmology \citep{2012JCAP...06..036S,2017JCAP...11..029Y}. One example is given in figure \ref{fig:GPR} for $E_{iso}$ and $\Delta t$ which are Gaussian distributed, other features, for example, the velocity of the system, the difference of the environment are considered as noise that also Gaussian distributed but with zero mean. The minimal of the GPR marginal likelihood gives the best fit, meanwhile it gives the noise level $\sim 8\%$. The curve fitted by GPR overlaps with the power-law line, which indicates the power-law assumption is appropriate.

\section{Discussion}

Many authors have attributed the origin of flares to the long activity of the GRB central engine: one possibility widely discussed is that the central engine produces a series of outgoing shells with a variety of independent Lorentz factors, the collisions between these shells occur over a wide time range, producing the prompt emission and the X-ray flares. \citep{1994ApJ...430L..93R,2005MNRAS.364L..42F,2007MNRAS.375L..46L,2007ApJ...658L..75G}. \citet{2016MNRAS.457L.108B}  consider the prompt emission and flares are emitted from the photospheres with different Lorentz factors. \citet{2005ApJ...630L.113K, 2006ApJ...636L..29P} propose the flares are produced by the continuous or discrete accretion from a surrounding disk. All of these models expect the prompt emission and the afterglow are similar entities but arise independently, no strong correlation should be found. \cite{2013NatPh...9..465W,2015ApJ...801...57G} applied the waiting time statistics, a common stochastic model was found to describe the prompt emission and the afterglow. This stochasticity in principle also shall lead to the independency of observational quantities between the prompt emission and the X-ray flare. But for our sample, although the prompt emission and the afterglow are two distinct phases, they are strictly correlated, there exists many correlations between the energy ($E_{iso}$) of the prompt emission and the energy ($E_{flare}$), time ($t_{peak}$), duration ($\Delta t$) of the flare. This discrepancy probably not come from the selection effect of our sample, since our criteria do not eliminate the randomness from the assumed central engine models. First all the criteria are apparently independent; second in the aforementioned papers, the characteristic parameters of criteria  have not been independently found any correlations with the characteristic variables of the models (Lorentz factor, accretion mass and etc); third, we don't specifically select some GRBs, our criteria make general divisions. Since the discrepancy is not from the selected data, we may state the long and random activity of central engine are not the solution for our sample. Furthermore, we are bold to propose that the observed stochasticity for a general selection of flares is not appropriately interpreted by different Lorentz factors or other characteristics within a specific mechanism, but flares could have many different origins, the statistical analysis adopting a mixture of flares with different origins fuzz up the intrinsic patterns and brings the stochastic appearance. The sample we selected may present one of the mechanisms, and the correlations indicate this mechanism of producing flares is highly related to the prompt emission, that the prompt emission and the X-ray flare may occur in a same sequence.

These correlations indicates the properties of central engine, moreover themselves have important applications: the prompt emission and afterglow are observed independently, the correlations we found bridge them, therefore, for GRBs without complete observation, our correlations can be used to infer the missing information; for GRBs with complete data, we can apply our correlations for the cosmology.

The launch of THESUS satellite\footnote{Official website: \url{http://www.isdc.unige.ch/theseus/}} will open our vision to the macroscopic early universe, as well as the microscopic details of the infra-red, X-ray and gamma-ray spectra. It will increase numerously the GRB sample for the statistics, also it will investigate deeply the commonalities and the differences between GRB components. In this paper, in order to understand the physics of GRBs, we adopted a novel perspective to study statistically the correlations between the prompt emission and the X-ray flare, the THESUS satellite perfectly suites our purpose, it will definitely promote our understanding to a higher stage.

\bibliographystyle{aa}

\begin{thebibliography}{}
\expandafter\ifx\csname natexlab\endcsname\relax\def\natexlab#1{#1}\fi

\bibitem[{{Ackermann} {et~al.}(2013){Ackermann}, {Ajello}, {Asano}, {Axelsson},
  {Baldini}, {Ballet}, {Barbiellini}, {Bastieri}, {Bechtol}, {Bellazzini},
  {Bhat}, {Bissaldi}, {Bloom}, {Bonamente}, {Bonnell}, {Bouvier}, {Brandt},
  {Bregeon}, {Brigida}, {Bruel}, {Buehler}, {Burgess}, {Buson}, {Byrne},
  {Caliandro}, {Cameron}, {Caraveo}, {Cecchi}, {Charles}, {Chaves},
  {Chekhtman}, {Chiang}, {Chiaro}, {Ciprini}, {Claus}, {Cohen-Tanugi},
  {Connaughton}, {Conrad}, {Cutini}, {D'Ammando}, {de Angelis}, {de Palma},
  {Dermer}, {Desiante}, {Digel}, {Dingus}, {Di Venere}, {Drell},
  {Drlica-Wagner}, {Dubois}, {Favuzzi}, {Ferrara}, {Fitzpatrick}, {Foley},
  {Franckowiak}, {Fukazawa}, {Fusco}, {Gargano}, {Gasparrini}, {Gehrels},
  {Germani}, {Giglietto}, {Giommi}, {Giordano}, {Giroletti}, {Glanzman},
  {Godfrey}, {Goldstein}, {Granot}, {Grenier}, {Grove}, {Gruber}, {Guiriec},
  {Hadasch}, {Hanabata}, {Hayashida}, {Horan}, {Hou}, {Hughes}, {Inoue},
  {Jackson}, {Jogler}, {J{\'o}hannesson}, {Johnson}, {Johnson}, {Kamae},
  {Kataoka}, {Kawano}, {Kippen}, {Kn{\"o}dlseder}, {Kocevski}, {Kouveliotou},
  {Kuss}, {Lande}, {Larsson}, {Latronico}, {Lee}, {Longo}, {Loparco},
  {Lovellette}, {Lubrano}, {Massaro}, {Mayer}, {Mazziotta}, {McBreen},
  {McEnery}, {McGlynn}, {Michelson}, {Mizuno}, {Moiseev}, {Monte}, {Monzani},
  {Moretti}, {Morselli}, {Murgia}, {Nemmen}, {Nuss}, {Nymark}, {Ohno},
  {Ohsugi}, {Omodei}, {Orienti}, {Orlando}, {Paciesas}, {Paneque}, {Panetta},
  {Pelassa}, {Perkins}, {Pesce-Rollins}, {Piron}, {Pivato}, {Porter}, {Preece},
  {Racusin}, {Rain{\`o}}, {Rando}, {Rau}, {Razzano}, {Razzaque}, {Reimer},
  {Reimer}, {Reposeur}, {Ritz}, {Romoli}, {Roth}, {Ryde}, {Saz Parkinson},
  {Schalk}, {Sgr{\`o}}, {Siskind}, {Sonbas}, {Spandre}, {Spinelli}, {Suson},
  {Tajima}, {Takahashi}, {Takeuchi}, {Tanaka}, {Thayer}, {Thayer}, {Thompson},
  {Tibaldo}, {Tierney}, {Tinivella}, {Torres}, {Tosti}, {Troja}, {Tronconi},
  {Usher}, {Vandenbroucke}, {van der Horst}, {Vasileiou}, {Vianello}, {Vitale},
  {von Kienlin}, {Winer}, {Wood}, {Wood}, {Xiong}, \&
  {Yang}}]{2013ApJS..209...11A}
{Ackermann}, M., {Ajello}, M., {Asano}, K., {et~al.} 2013, ApJS, 209, 11

\bibitem[{{Beniamini} \& {Kumar}(2016)}]{2016MNRAS.457L.108B}
{Beniamini}, P., \& {Kumar}, P. 2016, MNRAS, 457, L108

\bibitem[{{Bloom} {et~al.}(2001){Bloom}, {Frail}, \&
  {Sari}}]{2001AJ....121.2879B}
{Bloom}, J.~S., {Frail}, D.~A., \& {Sari}, R. 2001, AJ, 121, 2879

\bibitem[{Chincarini {et~al.}(2007)Chincarini, Moretti, Romano, Falcone,
  Morris, Racusin, Campana, Covino, Guidorzi, Tagliaferri, Burrows, Pagani,
  Stroh, Grupe, Capalbi, Cusumano, Gehrels, Giommi, La~Parola, Mangano, Mineo,
  Nousek, O'Brien, Page, Perri, Troja, Willingale, \&
  Zhang}]{2007ApJ...671.1903C}
Chincarini, G., Moretti, A., Romano, P., {et~al.} 2007, ApJ, 671, 1903

\bibitem[{Curran {et~al.}(2008)Curran, Starling, O'Brien, Godet, van~der Horst,
  \& Wijers}]{2008A&A...487..533C}
Curran, P.~A., Starling, R. L.~C., O'Brien, P.~T., {et~al.} 2008, A\&A, 487, 533--538

\bibitem[{{Dezalay} {et~al.}(1992){Dezalay}, {Barat}, {Talon}, {Syunyaev},
  {Terekhov}, \& {Kuznetsov}}]{1992AIPC..265..304D}
{Dezalay}, J.-P., {Barat}, C., {Talon}, R., {et~al.} 1992, in American
  Institute of Physics Conference Series, Vol. 265, American Institute of
  Physics Conference Series, ed. W.~S. {Paciesas} \& G.~J. {Fishman}, 304--309

\bibitem[{{Evans} {et~al.}(2007){Evans}, {Beardmore}, {Page}, {Tyler},
  {Osborne}, {Goad}, {O'Brien}, {Vetere}, {Racusin}, {Morris}, {Burrows},
  {Capalbi}, {Perri}, {Gehrels}, \& {Romano}}]{2007A&A...469..379E}
{Evans}, P.~A., {Beardmore}, A.~P., {Page}, K.~L., {et~al.} 2007, A\&A, 469,
  379

\bibitem[{{Evans} {et~al.}(2009){Evans}, {Beardmore}, {Page}, {Osborne},
  {O'Brien}, {Willingale}, {Starling}, {Burrows}, {Godet}, {Vetere}, {Racusin},
  {Goad}, {Wiersema}, {Angelini}, {Capalbi}, {Chincarini}, {Gehrels}, {Kennea},
  {Margutti}, {Morris}, {Mountford}, {Pagani}, {Perri}, {Romano}, \&
  {Tanvir}}]{2009MNRAS.397.1177E}
---. 2009, MNRAS, 397, 1177

\bibitem[{{Fan, Y Z} \& {Wei, D M}(2005)}]{2005MNRAS.364L..42F}
{Fan, Y Z}, \& {Wei, D M}. 2005, MNRAS, 364, L42

\bibitem[{Ghisellini {et~al.}(2007)Ghisellini, Ghirlanda, Nava, \&
  Firmani}]{2007ApJ...658L..75G}
Ghisellini, G., Ghirlanda, G., Nava, L., \& Firmani, C. 2007, ApJ, 658, L75

\bibitem[{Guidorzi {et~al.}(2015)Guidorzi, Dichiara, Frontera, Margutti,
  Baldeschi, \& Amati}]{2015ApJ...801...57G}
Guidorzi, C., Dichiara, S., Frontera, F., {et~al.} 2015, ApJ, 801, 57

\bibitem[{King {et~al.}(2005)King, O'Brien, Goad, Osborne, Olsson, \&
  Page}]{2005ApJ...630L.113K}
King, A., O'Brien, P.~T., Goad, M.~R., {et~al.} 2005, ApJ, 630, L113

\bibitem[{{Kouveliotou} {et~al.}(1993){Kouveliotou}, {Meegan}, {Fishman},
  {Bhat}, {Briggs}, {Koshut}, {Paciesas}, \& {Pendleton}}]{1993ApJ...413L.101K}
{Kouveliotou}, C., {Meegan}, C.~A., {Fishman}, G.~J., {et~al.} 1993, ApJL,
  413, L101

\bibitem[{Lazzati \& Perna(2007)}]{2007MNRAS.375L..46L}
Lazzati, D., \& Perna, R. 2007, MNRAS, 375, L46

\bibitem[{{Lien} {et~al.}(2014){Lien}, {Barthelmy}, {Chester}, {D'Elia},
  {Malesani}, {Maselli}, {Page}, {Palmer}, \& {Siegel}}]{2014GCN..15784...1L}
{Lien}, A.~Y., {Barthelmy}, S.~D., {Chester}, M.~M., {et~al.} 2014, GRB
  Coordinates Network, 15784

\bibitem[{{Malesani} {et~al.}(2014){Malesani}, {Xu}, {Fynbo}, {de Ugarte
  Postigo}, {Schulze}, {Finoguenov}, {Jakobsson}, {Melandri}, \&
  {Cucchiara}}]{2014GCN..15800...1M}
{Malesani}, D., {Xu}, D., {Fynbo}, J.~P.~U., {et~al.} 2014, GRB Coordinates
  Network, 15800

\bibitem[{Perna {et~al.}(2006)Perna, Armitage, \& Zhang}]{2006ApJ...636L..29P}
Perna, R., Armitage, P.~J., \& Zhang, B. 2006, ApJ, 636,
  L29

\bibitem[{Rees \& M{\'e}sz{\'a}ros(1994)}]{1994ApJ...430L..93R}
Rees, M.~J., \& M{\'e}sz{\'a}ros, P. 1994, ApJ, 430, L93--96

\bibitem[{Ruffini {et~al.}(2015)Ruffini, Aimuratov, Bianco, Enderli, Kovacevic,
  Moradi, Muccino, Penacchioni, Pisani, Rueda, \& Wang}]{2015IJMPA..3045023R}
Ruffini, R., Aimuratov, Y., Bianco, C.~L., {et~al.} 2015, IJMPA, 30, 1545023

\bibitem[{Ruffini {et~al.}(2016)Ruffini, Rueda, Muccino, Pisani, Wang, Becerra,
  Kovacevic, Oliveira, Aimuratov, Bianco, \& Moradi}]{2016ApJ...832..136R}
Ruffini, R., Rueda, J.~A., Muccino, M., {et~al.} 2016, ApJ, 832, 2

\bibitem[{{Ruffini} {et~al.}(2018){Ruffini}, {Wang}, {Aimuratov}, {Barres de
  Almeida}, {Becerra}, {Bianco}, {Chen}, {Karlica}, {Kovacevic}, {Li}, {Melon
  Fuksman}, {Moradi}, {Muccino}, {Penacchioni}, {Pisani}, {Primorac}, {Rueda},
  {Shakeri}, {Vereshchagin}, \& {Xue}}]{2018ApJ...852...53R}
{Ruffini}, R., {Wang}, Y., {Aimuratov}, Y., {et~al.} 2018, ApJ, 852, 53

\bibitem[{{Sakamoto} {et~al.}(2014){Sakamoto}, {Barthelmy}, {Baumgartner},
  {Cummings}, {Gehrels}, {Krimm}, {Lien}, {Markwardt}, {Palmer}, {Stamatikos},
  {Tueller}, \& {Ukwatta}}]{2014GCN..15805...1S}
{Sakamoto}, T., {Barthelmy}, S.~D., {Baumgartner}, W.~H., {et~al.} 2014, GRB
  Coordinates Network, 15805

\bibitem[{{Seikel} {et~al.}(2012){Seikel}, {Clarkson}, \&
  {Smith}}]{2012JCAP...06..036S}
{Seikel}, M., {Clarkson}, C., \& {Smith}, M. 2012, JCAP, 6, 036

\bibitem[{Swenson {et~al.}(2013)Swenson, Roming, de~Pasquale, \&
  Oates}]{2013ApJ...774....2S}
Swenson, C.~A., Roming, P. W.~A., de~Pasquale, M., \& Oates, S.~R. 2013, ApJ, 774, 2

\bibitem[{Wang \& Dai(2013)}]{2013NatPh...9..465W}
Wang, F.~Y., \& Dai, Z.~G. 2013, Nature Physics, 9, 465

\bibitem[{{Yennapureddy} \& {Melia}(2017)}]{2017JCAP...11..029Y}
{Yennapureddy}, M.~K., \& {Melia}, F. 2017, JCAP, 11, 029

\bibitem[{{Zhang} {et~al.}(2006){Zhang, B}, {Fan, Y Z}, {Dyks,
  Jaroslaw}, {Kobayashi, Shiho}, {Meszaros, Peter}, {Burrows, David N},
  {Nousek, John A}, \& {Gehrels, Neil}}]{2006ApJ...642..354Z}
{Zhang, B.}, {Fan, Y Z.}, {Dyks, J.}, {et~al.} 2006, ApJ, 642, 354

\end{thebibliography}

\end{document}